%% file: martellotti.tex
\documentclass[12pt]{article}
\usepackage{graphicx}
\usepackage{subfigure}
\newcommand\pubnumber{SNSN-323-63}
\newcommand\pubdate{\today}

\def\napoli{INFN, Laboratori Nazionali di Frascati (RM), Italy}
\def\support{\footnote{On behalf of the NA62 Collaboration:
G.~Aglieri Rinella, R.~Aliberti, F.~Ambrosino, B.~Angelucci, A.~Antonelli, 
G.~Anzivino, R.~Arcidiacono, I.~Azhinenko, 
S.~Balev, M.~Barbanera
% ...
}}

\def\Title#1{\begin{center} {\Large #1 } \end{center}}
\def\Author#1{\begin{center}{ \sc #1} \end{center}}
\def\Address#1{\begin{center}{ \it #1} \end{center}}

\newcommand\pubblock{\rightline{\begin{tabular}{l} \pubnumber\\
         \pubdate  \end{tabular}}}
\newenvironment{Abstract}{\begin{quotation}  }{\end{quotation}}
\newenvironment{Presented}{\begin{quotation} \begin{center} 
             PRESENTED AT\end{center}\bigskip 
      \begin{center}\begin{large}}{\end{large}\end{center} \end{quotation}}

\input econfmacros.tex

\begin{document}
\begin{titlepage}
\pubblock

\vfill
\Title{The NA62 Experiment at CERN}
\vfill
\Author{Silvia Martellotti\support}
\Address{\napoli}
\vfill
\begin{Abstract}
The main physics goal of the NA62 experiment at CERN is to precisely measure the branching ratio of the kaon rare decay $K^+\rightarrow \pi^+ \nu \bar\nu$. This decay is strongly suppressed in the Standard Model and its branching ratio is theoretically calculated with high accuracy. The NA62 experiment is designed to measure this decay rate with an uncertainty
better than 10\%. 
The measurement can be a good probe of new physics phenomena, which can
alter the SM decay rate. The NA62 experiment has been successfully launched in October 2014. In this document, after an introduction to the theoretical framework, the NA62 experimental setup is described and a first look at the pilot run data is reported.
\end{Abstract}
\vfill
\begin{Presented}
CIPANP.
Twelfth Conference on the Intersections of Particle and Nuclear Physics.
Vail (CO), USA ,  May 19--24, 2015
\end{Presented}
\vfill
\end{titlepage}
\def\thefootnote{\fnsymbol{footnote}}
\setcounter{footnote}{0}

\section{Introduction}

The NA62 (62nd experiment in the CERN North Area) \cite{NA62} is a fixed target experiment at CERN operating on the 400GeV/c proton beam supplied by the CERN Super-Proton-Synchrotron (SPS) facility. The goal of the experiment is to test the Standard Model (SM) in a kaon rare decay. A $K^+$ beam with momentum $p_{K^+} = 75$ GeV/c is selected from the proton beam impinging on a Beryllium target. The NA62 sub-detectors are located along the trajectory of the $K^+$ beam and they aim to identify kaon decay particles and to measure their momentum and energy. The NA62 experiment had a first pilot run in October-November 2014. Data collected in the pilot run are used to study the detector performance and to validate the analysis method. The NA62 experiment plans to collect data each year until 2018, when the second long shutdown of the Large Hadron Collider (LHC) machine is foreseen.

\section{Theoretical framework}

The $K\rightarrow \pi \nu \bar\nu$ decays are flavor-changing neutral current (FCNC) processes that
probe the $s\rightarrow d \nu \bar\nu$ transition via the Z-penguin and the box diagram shown in Figure \ref{fig:diag}. They are highly GIM suppressed and their Standard Model (SM) rates are very small.
For several reasons, the SM calculation for their branching ratios (BRs) is particularly clean \cite{teoria}:
\begin{figure}[htb]
\centering
\includegraphics[width=0.7\textwidth]{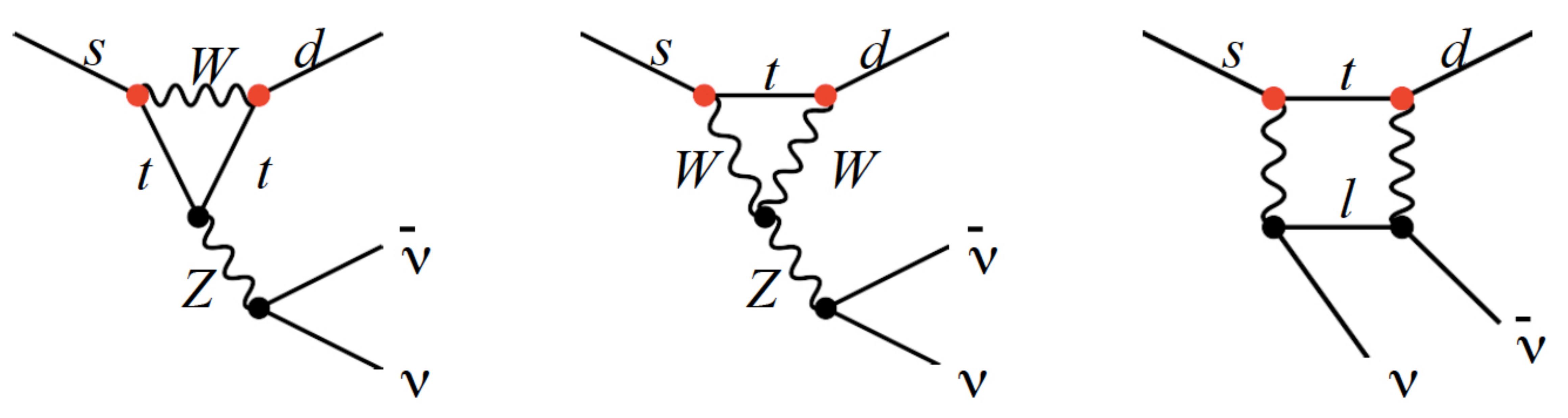}
\caption{Z-penguin diagrams and box diagram contributing to the process $K\rightarrow \pi \nu \bar\nu$.}
\label{fig:diag}
\end{figure}

\begin{itemize}
\item The loop amplitudes are dominated by the top-quark contributions. The neutral
decay violates CP; its amplitude involves the top-quark contribution only.
Small corrections to the amplitudes from the lighter quarks come into play for
the charged channel.
\item The hadronic matrix element for these decays can be obtained from the precise
experimental measurement of the $K_{e3}$ rate.
\item There are no long-distance contributions from processes with intermediate photons.
\end{itemize}

In the SM, $BR(K_L \rightarrow \pi^0 \nu \bar \nu)= (2.43\pm0.39\pm0.06) \times 10^{-11}$ and 
$BR(K^+ \rightarrow \pi^+ \nu \bar \nu)=(7.81\pm0.75\pm0.29) \times 10^{-11}$ \cite{BR_theory}. 
The uncertainties listed first derive from the input parameters.
The smaller uncertainties listed second demonstrate the size of the intrinsic
theoretical uncertainties. Because of the corrections from lighter-quark contributions,
these are slightly larger for the charged channel.
Because the SM rates are small and predicted very precisely, the BRs for these
decays are sensitive probes for new physics. In evaluating the rates for the
FCNC kaon decays, the different terms of the operator product expansion are differently
sensitive to modifications from a given new-physics scenario. If $BR(K_L \rightarrow \pi^0 \nu \bar \nu)$
and $BR(K^+ \rightarrow \pi^+ \nu \bar \nu)$ are ultimately both measured, and one or both BRs is found
to differ from its SM value, it may be possible to characterize the physical mechanism
involved [3], e.g., a mechanism with minimal flavor violation \cite{MFV}, manifestations
of supersymmetry \cite{superSim}, a fourth generation of fermions \cite{4gen}, Higgs compositeness as
in the littlest Higgs model \cite{Higgs}, or an extra-dimensional mechanism such as in the
Randall-Sundrum model \cite{RanSun}.

The most precise experimental result for the $BR(K^+ \rightarrow \pi^+ \nu \bar \nu)$ has been obtained
by the dedicated experiments E787 and E949 at the Brookhaven National Laboratory \cite{BNL} and is based on a total of 7 events using a decay-at-rest rechnique. The combined measurement is $BR(K^+ \rightarrow \pi^+ \nu \bar \nu)=(17.3^{+11.5}_{-10.5}) \times 10^{-11}$.

\section{Experimental requirements}

NA62 uses a decay-in-flight technique, differently from the kaon decay at rest approach which
was at the basis of the previous experiments. 

The presence of two undetectable neutrinos in the final state reduces the signal signature to one high momentum charged track with nothing else, which has to be discriminated against background coming from all other kaon decays.
The high momentum of the incoming beam (75 GeV/c) improves the background rejection and sets the longitudinal scale of the experiment. 
To achieve the required background suppression different principles have to be combined and the resulting requirements are outlined here:
\begin{itemize}
\item High intensity and good timing: a high intensity kaon beam is essential in order to reach sensitivity to a branching ratio of $\sim10^{-10}$. The incoming secondary beam from the SPS provides a particle rate of 750 MHz, containing about 6\% of kaons delivering roughly $45 \cdot 10^6$ decays in the fiducial region per spill. Precise timing (in the range 100 - 150ps) of the $K^+$ and the $\pi^+$ allows precise matching of the particles in the decay. The time resolution is essential to keep wrong associations below 1\%.

\item Low-mass tracking: track position and momenta have to be measured with high accuracy in low mass detectors. They are essential because inelastic scattering of beam particles in the detector material can mimic an isolated $\pi^+$ appearing like a signal event and hence contribute to the background.

\item Hermetic vetoing for photons and muons: the kinematic rejection must be accompanied by direct vetoing for photons (in particular for the $K^+ \rightarrow \pi^+ \pi^0$ background) requiring a typical
inefficiency of $10^{-4}$ for high energetic photons. About two third of $K^+$ decays contain muons in the final state, therefore a muon veto system is mandatory both in the trigger and off-line. 

\item Particle ID: several detectors must complement the event information with direct evidence on the particle species.

\end{itemize}

\section{NA62 experimental setup}

The experimental setup, shown in Figure \ref{fig:setup}, consists of a 100 m long beam line to select the appropriate secondary beam component produced by protons from the SPS CERN accelerator, followed by a 80 m long evacuated volume which defines the decay region.
Along the beam line, different detectors are distributed. From upstream:

\begin{figure}[htb]
\centering
\includegraphics[width=1.0\textwidth]{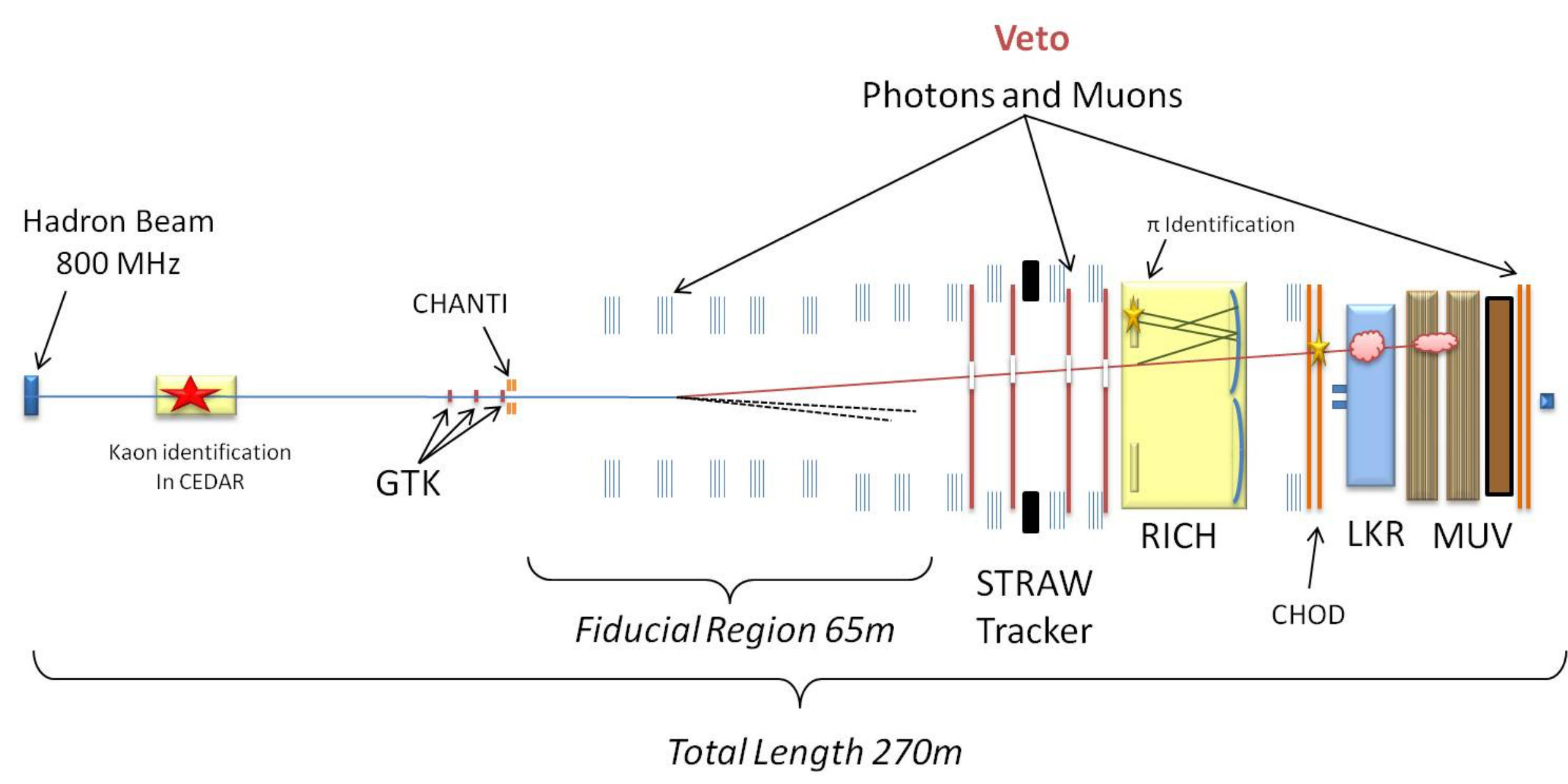}
\caption{Schematic view of the NA62 experiment showing the main sub-detectors (not to scale). The main elements for the detection of the $K^+$ decay products are located along a 170 m long region starting about $\sim 100$ m downstream of the beryllium target. Useful $K^+$ decays will be detected within a 65 m long fiducial decay region.}
\label{fig:setup}
\end{figure}

\begin{itemize}
\item{\bfseries Cedar and KTAG}:
The KTAG identifies the $K^+$ component in the beam with respect to the other beam particles by employing an upgraded differential Cherenkov counter (CEDAR).
The CEDAR filled with nitrogen gas is placed in the incoming beam to positively identify the kaon
component in the high rate environment. It is designed to identify particles of a specific mass by
making the detector blind to the Cherenkov light produced by particles of different mass. The KTAG
upgrade of NA62 acts in the Cherenkov light detection stage: to cope with the expected 45 MHz
kaon rate, 384 photomultipliers (PMTs) grouped in 8 light boxes are placed behind the 8 annular
slits. A preliminary analysis of 2014 data shows a kaon-identification efficiency better than 95\% and a time resolution below 100 ps.

\item{\bfseries Beam tracker GTK}:
The GigaTracker (GTK) comprises three stations measuring time, direction and momentum of the beam particles before entering the decay region.
The three stations are placed along the beam line just before the fiducial region. Each GTK station is a hybrid silicon micro-pixel detector with a total number of 18000 $300\times300$ $\mu$m$^2$ pixels grouped in 10 read-out chips. The expected performances on the track measurement are a 0.4\% resolution for the momentum, 16 mrad for the direction and 200 ps for the timing. One chip per station has been commissioned during the 2014 run.

\item{\bfseries Guard-Ring detector CHANTI}:
this detector is a guard-ring counter following the last GTK station to veto beam particles which scatter inelastically in the material of the third GTK station.
It consists of six stations of x-y plastic scintillator bars surrounding the beam, covering
an angle between 0.034 and 1.38 rad. The expected efficiency in vetoing signal-like events is 99\%.
Preliminary 2014 data analysis shows a time resolution of the order of 1 ns.

\item{\bfseries Pion spectrometer STRAW}:
The STRAW Tracker measures the coordinates and momentum of secondary charged particles originating from the decay region. To minimise multiple scattering the chambers are built of ultra-light material and are installed inside the vacuum tank. Two straw chambers are located on each side of a large aperture dipole magnet, providing a vertical B-field of 0.36T.
Each chamber is equipped with 1792 straw tubes which are arranged in four ``views'' providing measurements of four coordinates $x, y, u, v$. The requirements for the detector performances are a relative momentum resolution of 1\%, a spatial resolution of 130 $\mu$m per coordinate and a very low track reconstruction inefficiency. The detector has been fully commissioned during the 2014 run.

\item{\bfseries Pion Timing CHOD}:
The CHOD is a Charged-particle HODoscope (inherited from NA48), covering the acceptance, to detect tracks with precise measurements of the arrival time and impact point, able to provide a fast signal used to drive the trigger. It is composed of two planes made of 64 plastic scintillators, one with vertical and one with horizontal slabs. The CHOD can provide the timing of charged decay products with a resolution of about 200 ps.

\item{\bfseries Particle ID detector RICH}:
The Ring Imaging CHerenkov detector (RICH) is situated downstream of the last straw chamber. It consists of a 17 m long radiator filled with Neon Gas at 1 atm. The Cherenkov light is reflected by a system of hexagonal spherical mirrors placed at the downstream end of the tank and detected by $2\times960$ photomultipliers located in the upstream end of the detector. It allow an additional rejection of the muon background. The $\mu$ suppression
factor is $10^{-2}$ between 15 and 35 GeV/c (the momentum range of interest), and a time resolution of 65 ps has been measured. 

\item{\bfseries Large-Angle Photon Vetoes LAV}:
The LAV system provides full coverage for decay photons with polar angles from 8.5 to 50 mrad,
with an inefficiency of $10^{-3} \div10^{-4}$ on photons down to 150 MeV. It consists of 12 stations, whose
diameter increases with distance from the target, containing 4 or 5 layers of azimuthally staggered
lead glass blocks, for a total of 2500 channels. The time resolution observed in 2014 run is about 1
ns.

\item{\bfseries Small-Angle Photon Vetoes IRC and SAC}:
A Small Angle Calorimeter (SAC) and an Intermediate Ring Calorimeter (IRC) give photon coverage below 1 mrad. Both detectors are made of consecutive lead and plastic scintillator plates. The inefficiency for 1 GeV photons is below $10^{-4}$. A time resolution of 3 ns was measured in the 2014 run.

\item{\bfseries Particle ID detector LKr}:
The LKr Calorimeter (inherited from NA48) measures precisely the electro-magnetic energy with the possibility
to discriminate between positrons and pions using their shower properties. 
It covers intermediate angle between 1 and 8.5 mrad. It is divided in 13248 cells and the electromagnetic
showers are fully detected through ionisation of low energy charged particles. The reconstructed
shower time resolution is 500 ps, and the space resolution of the order of 1 mm; the total inefficiency
for 10 GeV photons is below $10^{-5}$. The new readout has been validated during the 2014
run.

\item{\bfseries Particle ID detector Muon Veto (MUV)}:
NA62 uses a fast scintillator (MUV3) for direct muon vetoing and 2 sampling hadron calorimeters (MUV1+2) which measure the deposited hadron energy in the event, distinguishing hadron showers from muons. MUV1 and MUV2 modules are iron-scintillator sandwich calorimeters with scintillator strips alternately oriented in the horizontal and vertical directions. This system supplements and provides redundancy with respect to the RICH in the detection and rejection of muons. Only one of these was installed in 2014. The MUV3 is a vertical array of fast scintillator tiles placed after an iron wall and is used to detect nonshowering muons. Each tile is read by 2 PMTs to avoid the effect of Cherenkov photons on time resolution. Data from the 2014 run show a time resolution below 500 ps.

\end{itemize}

The experiment makes use of an integrated trigger
and data acqusition system with three trigger levels. The lowest level, level 0, is
implemented directly in the digital readout card for each detector subsystem. The
detector hits are resolved into quantities that are used in
trigger logic to decide which events will be read out for level 1. Level 0 will process
about 10 MHz of ``primitive'' detector hits; about 1 MHz of events will be read out
for level 1. The level 1 trigger is implemented in software. 
It is the first asynchronous trigger level and will reduce the rate of
events seen by level 2 by an order of magnitude. The level 2 trigger is implemented
in the event builder running on the acquisition PC farm; it is the first trigger level
which makes use of the full event information. The O(100 kHz) of events input
to level 2 are reduced to a few kHz of events ultimately written to disk.

\section{Analysis strategy}

NA62 detectors are designed to measure the $K^+$ and the secondary particles from kaon decays occurring
in the decay volume.
The signature of the signal is one track in the final state matched with one $K^+$ track in the beam.
The measurable quantities are $\vec p_{K^+}$ (the kaon momentum-vector), $\vec p_{\pi^+}$ (the pion momentum-vector) and the angle $\theta_{\pi K}$ (angle between the pion and the kaon).

The main kinematic variable to discriminate the signal from the overwhelming background is the squared
missing mass $m^2_{miss}$. It is defined for the events, when only one charged track from Kaon decay is detected assuming the track belongs to a charged pion,
\begin{equation}
m^2_{miss} = (p_{K^+}-p_{\pi^+}).
\end{equation}

\begin{figure}
 \centering
 \subfigure
   {\includegraphics[width=7cm]{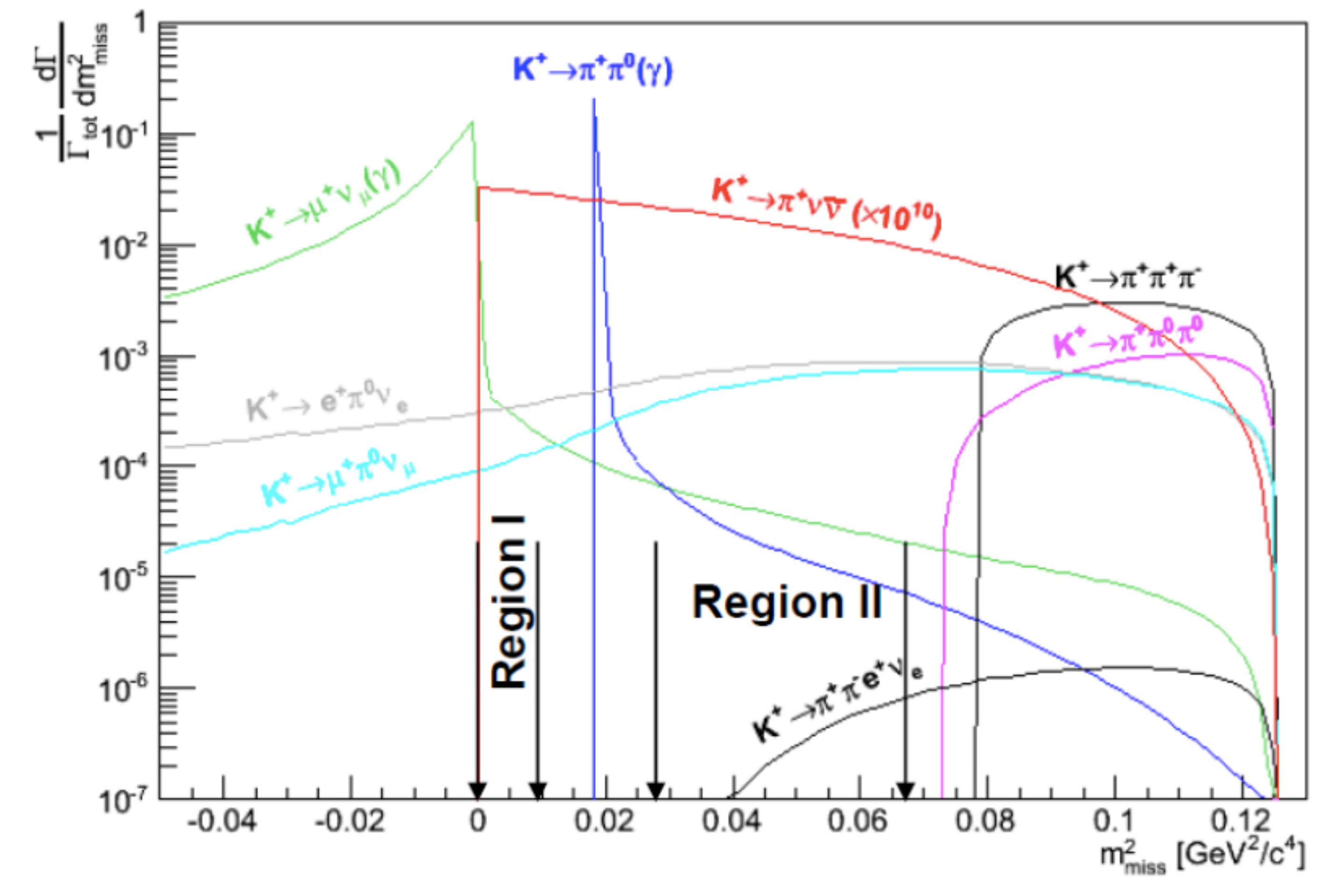}}
 \hspace{5mm}
 \subfigure
   {\includegraphics[width=7cm]{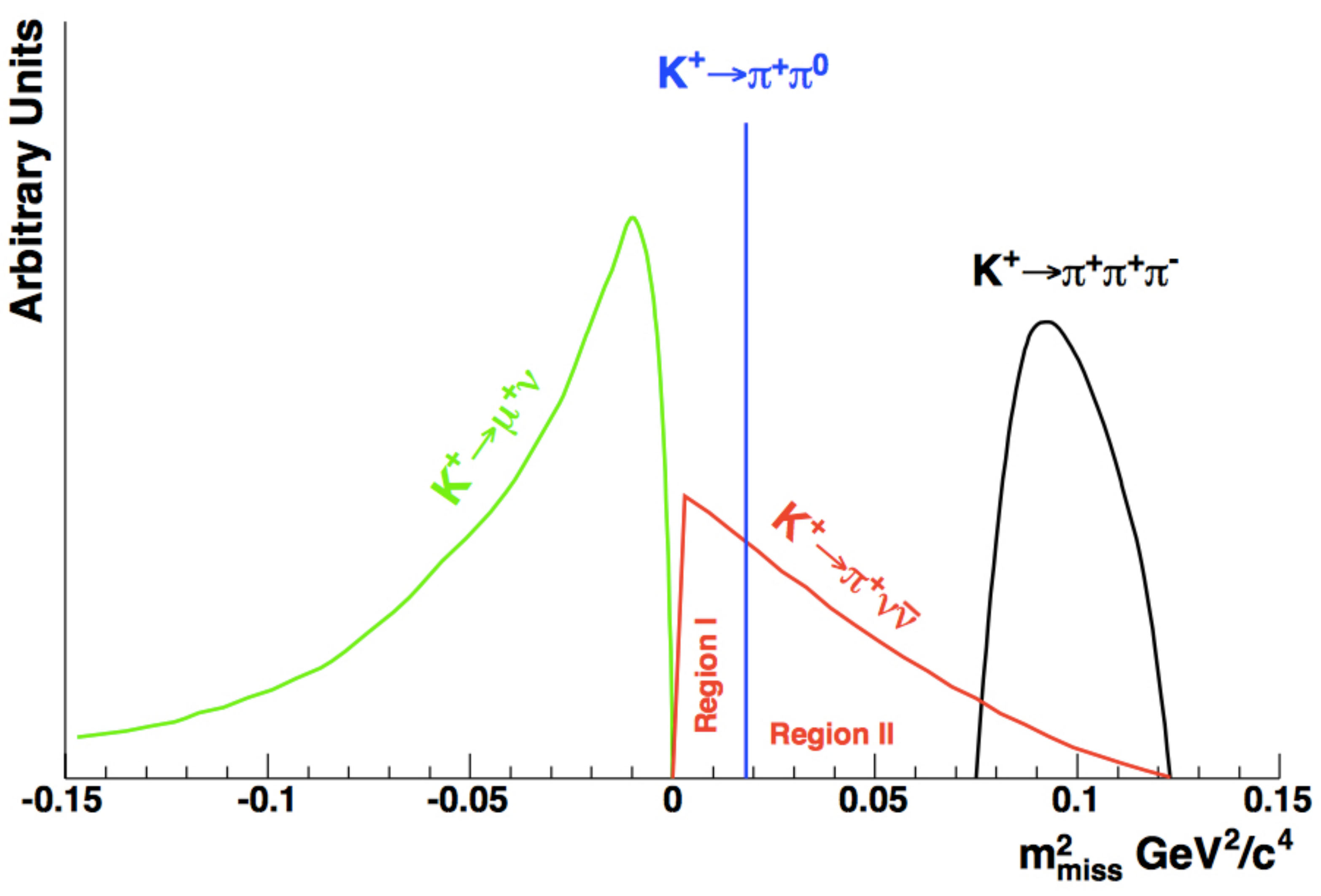}}
 \caption{Left: Distribution of $m^2_{miss}$ for the main kaon decays and the signal, with the hypothesis that the charged decay product is a pion. The two signal regions, $[0,0.01]GeV^2/c^4$ and
$[0.026, 0.068]GeV^2/c^4$, are indicated. Right: Distribution of $m^2_{miss}$ for the signal and the main background sources after kinematic cuts.}
\label{fig:MissMassT}
 \end{figure}

Figure \ref{fig:MissMassT} (left) shows the $m^2_{miss}$ distribution in log scale for the main decay channels and the
 $K^+ \rightarrow \pi^+ \nu \bar \nu$ decay (red curve). The signal decay differential width is magnified by $10^{10}$. 
Two signal regions (I and II), on each side of the $K^+ \rightarrow \pi^+\pi^0$ peak, are chosen, where more than 90\% of the main $K^+$ decays do not contribute. The kinematics background suppression is of the order of $10^5$. The further suppression of background at the order of $10^7$ is done by means of the detector related cuts (timing, particle identification, muon and photon veto).
 
Table \ref{tab:blood} shows the expected number of signal and background events per year of data taking estimated from MC simulation. Given this expectation, NA62 will be able to reach better than 10\% precision
in a measurement of $BR(K^+ \rightarrow \pi^+ \nu \bar \nu)$. The main background remains the 
$K^+\rightarrow \pi^+\pi^0(\gamma)$ decay, where two photons from $\pi^0$ decay (and FSR photon) are not detected and the signal event topology is mimicked. To suppress this background very good photon veto ability is required. The aimed $\pi^0$ suppression factor is $10^8$. The second important background comes from the $K^+ \rightarrow \mu^+ \nu_\mu (\gamma)$ decay. The designed overall suppression factor for muons is of the order of $10^5$.

\begin{table}[ht]
\begin{center}
\begin{tabular}{|l|c|}  
\hline
\small{Decay}  & \small{event/year}  \\ 
\hline
\small{$K^+ \rightarrow \pi^+ \nu\bar \nu$}  &   \small{45}  \\
\hline
\small{$K^+ \rightarrow \pi^+ \pi^0$}  &   \small{5}  \\ 
 \small{$K^+ \rightarrow \pi^+ \pi^+ \pi^-$}  &  \small{1}  \\
 \small{$K^+ \rightarrow \pi^+\pi^- e^+ \nu_e$}  &   \small{$<1$}  \\
\small{ $K^+ \rightarrow \mu^+ \nu_\mu \gamma$}  &   \small{1.5}  \\
\small{ other rare decays} &   \small{0.5}\\
 \hline
\small{ Total background} & \small{$<10$} \\
 \hline
\end{tabular}
\caption{Expected signal and background from $K^+$ decays estimated from NA62 sensitivity studies.}
\label{tab:blood}
\end{center}
\end{table}

In the analysis, the $\pi^+$ momentum will be required to be less than 35 GeV/c. In this way, the momentum of the $\pi^0$ amounts to at least 40 GeV/c. Such a large energy deposit can hardly be missed in the calorimeters. In addition only events with the $K^+$ vertex reconstructed in the fiducial volume are selected. The $m^2_{miss}$ distribution for signal and the three main background sources after the kinematic cuts is shown in Figure \ref{fig:MissMassT} (right).

\section{2014 pilot run}

Data recorded during the pilot run in 2014 are very successfully used to understand the performance of the
NA62 sub-detectors. However, only a small part of it was recorded with stable run conditions and is useful for preliminary physics studies. The beam intensity was 5\% of the nominal one and two triggers have been used to collect a minimum bias sample and a $\pi\nu\nu$-like without photon rejection sample. Only about 1\% of the available data have been studied. The experiment was at a preliminary stage: data were not yet readout from GTK and the analysis used the mean nominal direction and momentum of the kaon beam; LAV information was not yet used for photon veto; only preliminary detector time alignment and energy calibration were used.

\begin{figure}[htb]
\centering
\includegraphics[width=1.0\textwidth]{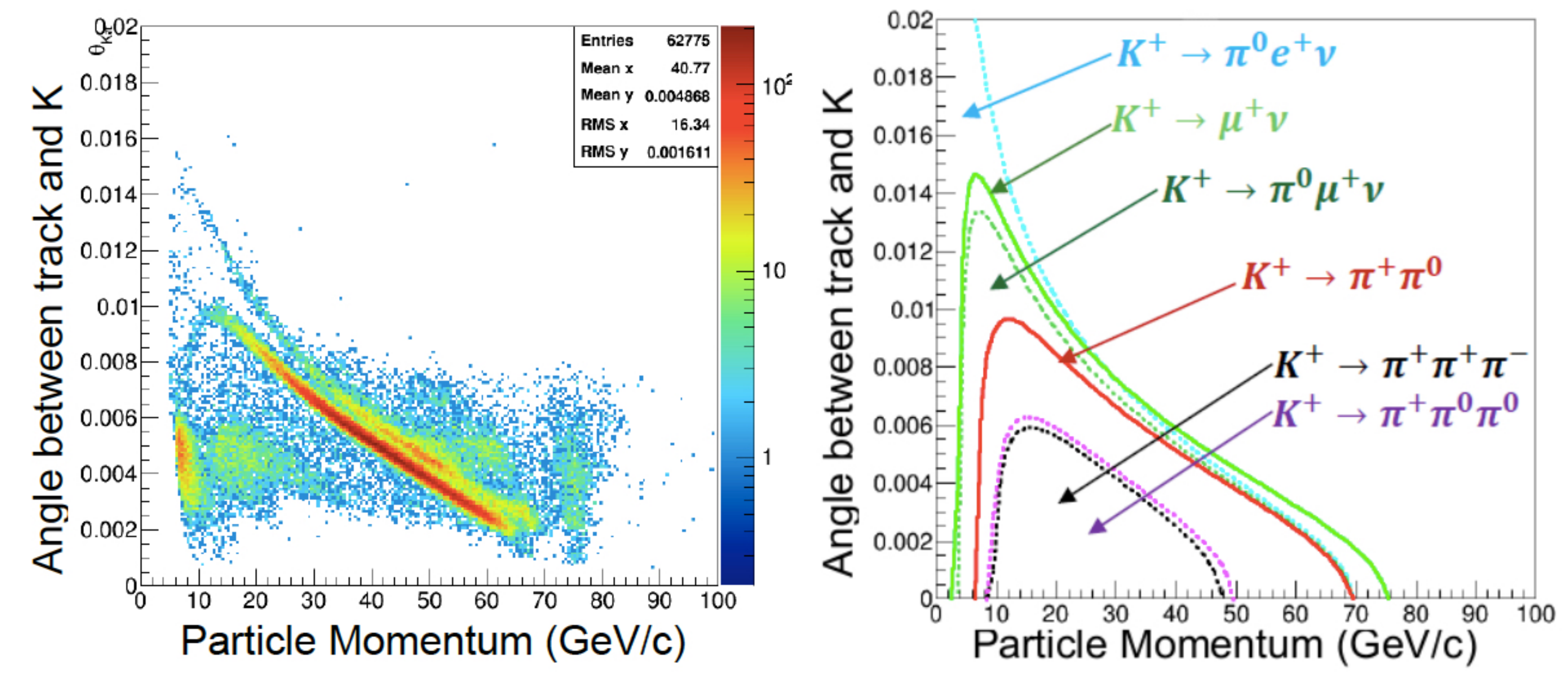}
\caption{Angle between the reconstructed track and the nominal kaon direction as a
function of particle momentum [GeV /c] in data (left) and as expected from pure kinematics (right).}
\label{fig:plotRes}
\end{figure}

\begin{figure}[htb]
\centering
\includegraphics[width=0.5\textwidth]{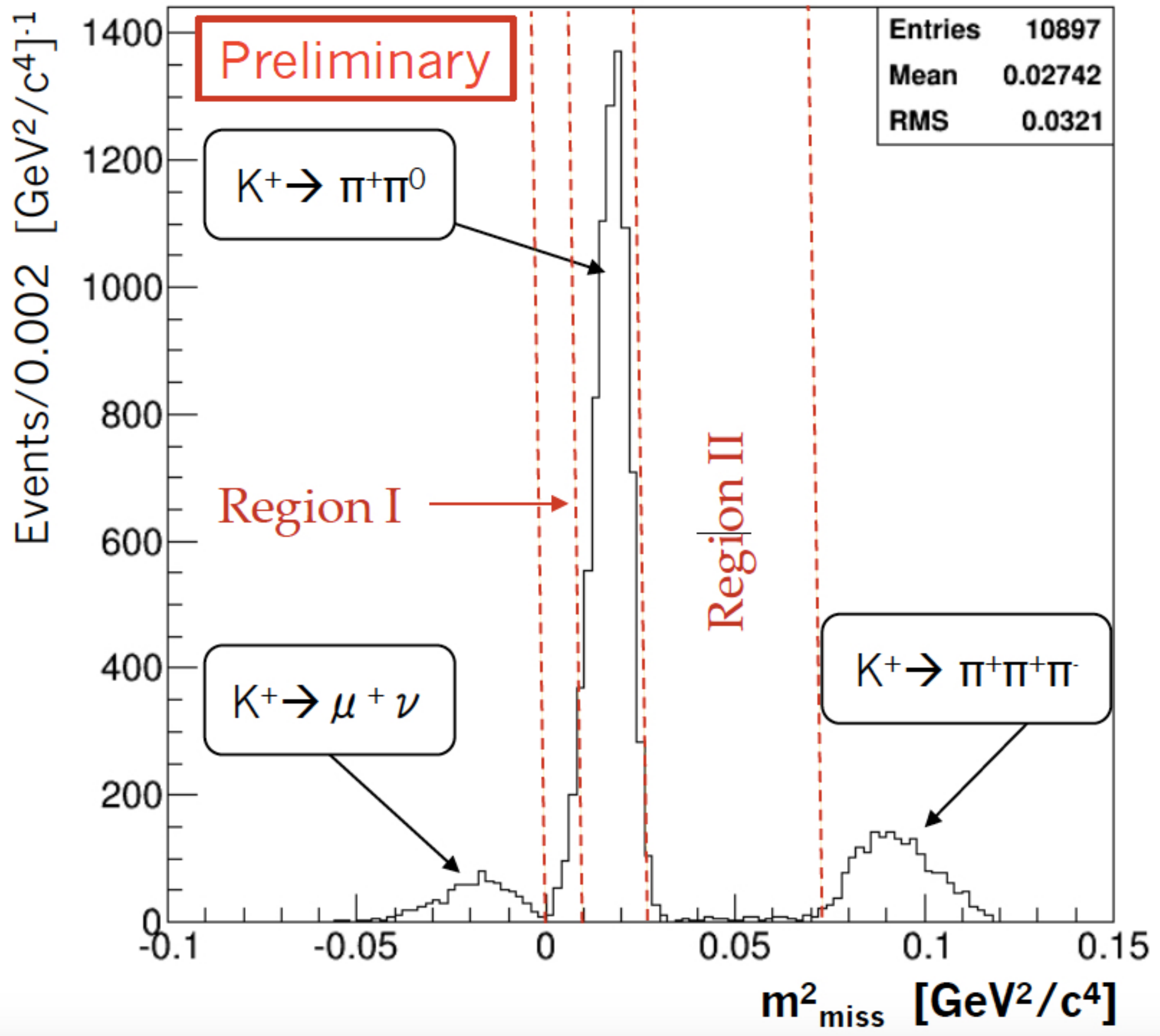}
\caption{the $m^2_{miss}$ distribution after the signal event selection. Two
regions for the final selection of signal events are indicated by the
vertical red-dashed lines. The kaon decays contributing to the distribution
are indicated.}
\label{fig:MissExp}
\end{figure}

The distribution of the angle between the kaon and the secondary track as a function of the track
momentum is shown in Figure \ref{fig:plotRes} (left), for events with only 1 track in the spectrometer and after requiring the presence of a kaon in the KTAG. A reasonable agreement can be observed if compared
with the expectations for the main kaon decays (right): the suppression of the $K^+\rightarrow \mu^+ \nu$
component is due to the muon rejection at trigger level.
The intersection between the track and the nominal beam direction is used to reconstruct the vertex
and suppresses the background from kaon interactions. 
After requiring the vertex position to be reconstructed in the fiducial decay region, 
and selecting the track momentum between 15 and 35 GeV/c, the squared missing mass
distribution shown in Figure \ref{fig:MissExp} is obtained.
The $K^+\rightarrow \mu^+ \nu$ background is
well suppressed with the help of RICH and muon veto system. The muon background suppression is expected
to be improved during the 2015 physics run with the help of MUV1. The three-track background, mainly
coming from the $K^+\rightarrow \pi^+ \pi^+ \pi^-$ decay is also well suppressed and does not contribute to the signal selected region. The main background contribution comes from the $K^+\rightarrow \pi^+ \pi^0$ decay. This particular background dominates in the signal region after the final selection, especially in region I.
LAV information  is not yet exploited both at trigger level and analysis level, preventing an additional suppression of 
$K^+\rightarrow \pi^+ \pi^0$ while the use of mean beam momentum and direction, instead of event by event GTK information, worsens the $m^2_{miss}$ resolution.

\end{document}

%% file: econfmacros.tex
\def\beq{\begin{equation}}
\def\eeq#1{\label{#1}\end{equation}}
\def\eeqn{\end{equation}}

\def\beqa{\begin{eqnarray}}
\def\eeqa#1{\label{#1}\end{eqnarray}}
\def\eeqan{\end{eqnarray}}

\let\bar=\overbar

\def\Dslash{\not{\hbox{\kern-4pt $D$}}}
\def\dslash{\not{\hbox{\kern-2pt $\del$}}}

\def\msb{{\bar{\ssstyle M \kern -1pt S}}}




%% file: martellotti.bbl
\begin{thebibliography}{99}


\bibitem{NA62} na62.web.cern.ch/na62/Documents/TechnicalDesign.html, 2010.

\bibitem{teoria}
J. Brod, M. Gorbahn and E. Stamou, Phys. Rev. D 83, 034030 (2011)
[arXiv:1009.0947 [hep-ph]]

\bibitem{BR_theory}
D. M. Straub, arXiv:1012.3893 [hep-ph].

\bibitem{MFV}
T. Hurth, G. Isidori, J. F. Kamenik and F. Mescia, Nucl. Phys. B 808, 326
(2009) [arXiv:0807.5039 [hep-ph]].

\bibitem{superSim}
G. Isidori, F. Mescia, P. Paradisi, C. Smith and S. Trine, JHEP 0608, 064 (2006)
[hep-ph/0604074].

\bibitem{4gen}
A. J. Buras, B. Duling, T. Feldmann, T. Heidsieck and C. Promberger, JHEP
1009, 104 (2010) [arXiv:1006.5356 [hep-ph]].

\bibitem{Higgs}
M. Blanke, A. J. Buras, B. Duling, S. Recksiegel and C. Tarantino, Acta Phys.
Polon. B 41, 657 (2010) [arXiv:0906.5454 [hep-ph]].

\bibitem{RanSun}
M. Blanke, A. J. Buras, B. Duling, K. Gemmler and S. Gori, JHEP 0903, 108
(2009) [arXiv:0812.3803 [hep-ph]].

\bibitem{BNL}E949 COLLABORATION; ARTAMONOV, A.V. ET AL. New Measurement of the $K^+\rightarrow \pi^+ \nu \bar\nu$ branching ratio. Phys. Rev. Lett. 101 191802, 2008.


\end{thebibliography}
